\documentclass[twocolumn]{jpsj2} 

\usepackage{bm}

\title{Antiferromagnetic phase transition in four-layered high-$T_{\rm c}$ superconductors Ba$_2$Ca$_3$Cu$_4$O$_8$(F$_y$O$_{1-y}$)$_2$ with $T_{\rm c}=55-102$ K: $^{63}$Cu- and $^{19}$F-NMR studies}

\author{Sunao \textsc{Shimizu}$^{1}$\thanks{E-mail: shimizu@nmr.mp.es.osaka-u.ac.jp}, 
Hidekazu \textsc{Mukuda}$^{1}$, 
Yoshio \textsc{Kitaoka}$^{1}$, \\
Hijiri \textsc{Kito}$^{2}$,
Yasuharu \textsc{Kodama}$^{2}$,
Parasharam M. \textsc{Shirage}$^{2}$, 
and Akira \textsc{Iyo}$^{2}$}

\inst{$^{1}$Graduate School of Engineering Science, Osaka University, Toyonaka, Osaka 560-8531, Japan \\
$^{2}$National Institute of Advanced Industrial Science and Technology (AIST), Umezono, Tsukuba 305-8568, Japan \\}

\abst{We report on magnetic characteristics in four-layered high-$T_{\rm c}$ superconductors Ba$_2$Ca$_3$Cu$_4$O$_8$(F$_y$O$_{1-y}$)$_2$ with apical fluorine through $^{63}$Cu- and $^{19}$F-NMR measurements. The substitution of oxygen for fluorine at the apical site increases the carrier density ($N_{\rm h}$) and $T_{\rm c}$ from 55 K up to 102 K. The NMR measurements reveal that antiferromagnetic order, which can uniformly coexist with superconductivity, exists up to $N_{\rm h}$ $\simeq$ 0.15, which is somewhat smaller than $N_{\rm h}$ $\simeq$ 0.17 being the quantum critical point (QCP) for five-layered compounds. The fact that the QCP for the four-layered compounds moves to a region of lower carrier density than for five-layered ones ensures that the decrease in the number of CuO$_2$ layers makes an interlayer magnetic coupling weaker.}

\kword{high-T{\rm c} superconductivity, copper-oxide, antiferromagnetism, NMR, phase diagram, apical-fluorine}

\begin{document}
\maketitle

\section{Introduction}

Remarkably high superconducting transition temperature ($T_{\rm c}$) in the copper oxides has been realized in a multilayered structure of CuO$_2$ planes. $T_{\rm c}$ depends on the number of CuO$_2$ layers ($n$) in multilayered compounds with a maximum at $n=$ 3.\cite{Scott} In particular, the highest $T_{\rm c}$ was observed around 133 K in a Hg-based three-layered ($n=$3) copper oxide HgBa$_2$Ca$_2$Cu$_3$O$_y$ (Hg-1223).\cite{Schilling}  Copper oxides with more than three layers comprise inequivalent types of CuO$_2$ layers, an outer CuO$_2$ plane (OP) in a five-fold pyramidal coordination, and an inner plane (IP) in a four-fold square coordination (see Fig.\ref{fig:crystal,Tc} as example). Site-selective $^{63}$Cu-NMR studies have revealed that the local carrier density ($N_{\rm h}$) for the IP is smaller than that for the OP. These results in turn revealed an intimate relationship between antiferromagntism (AFM) and superconductivity (SC) inherent to CuO$_2$ layers into which mobile hole carriers were homogeneously doped.\cite{Tokunaga,Kotegawa2001,Kotegawa2004,Mukuda2006} 

The recent systematic Cu-NMR studies on five-layered ($n=5$) compounds have unraveled that AFM order, which can uniformly coexist with SC, is robust up to $N_{\rm h}$ $\simeq$ 0.17, a quantum critical point (QCP) where the AFM order collapses \cite{Mukuda2008}. This result significantly differs  from the well-established results for mono-layered ($n=1$) La$_{2-x}$Sr$_x$CuO$_4$ (LSCO) \cite{LSCO} and bi-layered ($n=2$) YBa$_2$Cu$_3$O$_{6+x}$ (YBCO),\cite{YBCO} in which the AFM order collapses completely by doping extremely small amount of holes of $N_{\rm h}\sim 0.02$ and 0.055, respectively. These results strongly suggest that the QCP of $n \leq 4$ moves to a region of lower carrier density than that of $n=5$. Therefore, it is likely that the interlayer magnetic coupling for an onset of AFM order is enhanced as a number of CuO$_2$ layers increases. In order to establish how the interlayer magnetic coupling affects the onset of AFM order, we deal with four-layered high-$T_{\rm c}$ superconductors Ba$_2$Ca$_3$Cu$_4$O$_8$(F$_y$O$_{1-y}$)$_2$ with apical fluorine. We note that all these compounds are in an underdoped state of hole doping regime.

Ba$_2$Ca$_3$Cu$_4$O$_8$(F$_y$O$_{1-y}$)$_2$ comprises a stack of four CuO$_2$ layers as illustrated in Fig.\ref{fig:crystal,Tc}(a). It is known as a new family of multilayered copper oxides with apical fluorine.\cite{Iyo_TcVsn,Iyo2,Iyo1}  Substitution of oxygen for apical fluorine, i.e. a decrease in nominal fluorine content ($y$) results in doping holes into CuO$_2$ layers, increasing $T_{\rm c}$ from 55 K at $y=1.0$ to 102 K at $y=0.6$ as indicated in Fig.\ref{fig:crystal,Tc}(b).\cite{Iyo2}  This system provides an opportunity to investigate the characteristics of CuO$_2$ layers over a wide range of carrier density, especially enabling us to focus on the interplay of SC and AFM in an underdoped region.\cite{Miller,Lake,Inaba,Himeda,Sidis,TKLee,Demler,Zhang,Lee}
We reported in the literature \cite{Shimizu} that {\it self-doping} occurred at $y=1.0$ to realize superconductivity. 
For a nominal content at $y=1.0$, if the apical site of OP were fully occupied by F$^{-1}$,  the formal Cu valence would be just +2 and hence a Mott insulator; however, it exhibits SC. The occurrence of the SC at $y=1.0$ was argued in terms of a {\it self-doping} model wherein charge carriers were transferred between IP and OP.\cite{Chen, OK, Shimizu} These discussions, however, were based on a simple assumption that all the apical sites were occupied by F$^{-1}$. Since then, a bi-layered apical-F compound Ba$_2$CaCu$_2$O$_4$F$_2$ has also been synthesized; it exhibits SC with $T_{\rm c}=$ 73 K.\cite{Iyo_unpublished}  The {\it self-doping} mechanism cannot apply to this compound because it has no IPs. We, therefore, have conducted NMR studies on the bi-layered system to confirm that the {\it self-doping} mechanism did not occur, and that all  multilayered compounds with the apical fluorine were doped by hole carriers irrespective of $y$. For $y=1.0$, it was anticipated that a possible replacement of O$^{-2}$ for F$^{-1}$ and/or excess oxygen in the BaF layers resulted in doping hole carriers into CuO$_2$ layers.\cite{Shimizu2} 

In this paper, we report systematic $^{63}$Cu- and $^{19}$F-NMR studies on Ba$_2$Ca$_3$Cu$_4$O$_8$(F$_y$O$_{1-y}$)$_2$ with $y=$ 0.6, 0.7, 0.8 and 1.0 as each nominal content. Measurements of $^{63}$Cu Knight shift ($^{63}K$) have revealed that hole carrier density ($N_{\rm h}$) increases progressively with decreasing $y$. The substitution of oxygen for fluorine at the apical site increases $N_{\rm h}$ and $T_{\rm c}$ from 55 K up to 102 K. 
The measurements of $^{63}$Cu-NMR spectra and nuclear-spin-lattice-relaxation rate of $^{19}$F-NMR ($^{19}(1/T_1)$) unravel that an AFM order, which can uniformly coexist with SC, exists up to $N_{\rm h}$ $\simeq$ 0.15 being a QCP for the four-layered compounds. From the fact that the QCP of the four-layered compounds moves to a region of lower carrier density than that of the five-layered ones, $N_{\rm h}$ $\simeq$ 0.17, it is ensured that the decrease in the number of CuO$_2$ layers makes an interlayer magnetic coupling weaker.

\section{Experimental}

\begin{figure}[tpb]
\begin{center}
\includegraphics[scale = 0.48]{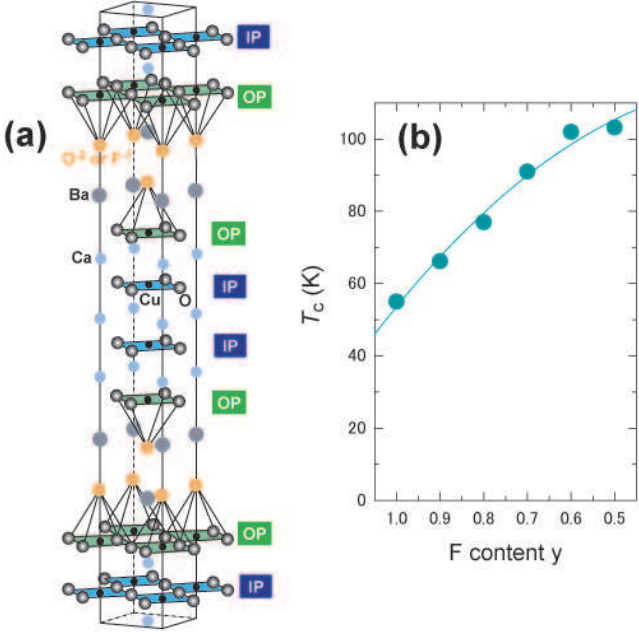}
\end{center}
\caption{\footnotesize (color online) (a) Crystal structure of Ba$_2$Ca$_3$Cu$_4$O$_8$(F$_y$O$_{1-y}$)$_2$. This four-layered system includes two crystallographically different CuO$_2$ planes, namely, IP and OP. (b) $T_{\rm c}$ vs. fluorine content $y$ \cite{Iyo2}. It shows that the substitution of oxygen for fluorine results in doping hole carriers into CuO$_2$ layers }
\label{fig:crystal,Tc}
\end{figure}

Polycrystalline powder samples of all multilayered systems used in this study were prepared by a high-pressure synthesis technique, as described elsewhere.\cite{Iyo_TcVsn,Iyo1,Iyo2}  Powder X-ray diffraction measurements indicate that these compounds almost entirely comprise a single phase, and that the $a$-axis length continually changes with the nominal fluorine content $y$ \cite{Iyo2}. The $T_{\rm c}$ was uniquely determined by a sharp onset of diamagnetism using a dc SQUID magnetometer as summarized in Table \ref{t:ggg}. For NMR measurements, the powder samples were aligned along the $c$-axis at an external field ($H$) of $\sim$ 16 T, and fixed using stycast 1266 epoxy. The NMR experiments were performed by the conventional spin-echo method in a temperature ($T$) range of 1.5 $-$ 300 K with $H$ perpendicular or parallel to the $c$-axis.
 
\section{Results and Discussions}

\subsection{Knight shift and local carrier density}
\begin{table}[b]
\caption{\footnotesize Lists of $T_{\rm c}$ and $N_{\rm h}$ at the OP and IP of Ba$_2$Ca$_3$Cu$_4$O$_8$(F$_y$O$_{1-y}$)$_2$. $N_{\rm h}$ is estimated from the $K_{\rm s}(T)$ at room temperature (see text). $\overline{N_{\rm h}}$ is an average of the carrier density, defined as ($N_{\rm h}$(OP) + $N_{\rm h}$(IP))/2. Note that $y$ is nominal fluorine contents.}
 \begin{center}
  \begin{tabular}{ccccc} \hline\hline
                     & $y=0.6$ & $y=0.7$ & $y=0.8$ & $y=1.0$ \\ \hline
    $T_{\rm c}$      & 102 K       & 91 K        & 77 K        & 55 K        \\
    $N_{\rm h}$ [OP] & + 0.207       & + 0.189       & + 0.167       & + 0.148       \\
    $N_{\rm h}$ [IP] & + 0.165       & + 0.150       & + 0.144       & $^*$(+ 0.132)       \\
    $\overline{N_{\rm h}}$ & + 0.186 & + 0.170 & + 0.156 & $^*$(+ 0.140)                    \\ \hline\hline \\
    
  \end{tabular}
 \end{center}
\footnotesize{$*$) $\overline{N_{\rm h}}=0.140$ at $y$=1.0 is estimated at $N_{\rm h}$(OP)=0.148 on a linear line in the plot of $N_{\rm h}$(OP) vs $\overline{N_{\rm h}}$ (see Fig.\ref{fig:av}).  A linear extrapolation in the  plot in $N_{\rm h}$(IP) vs. $\overline{N_{\rm h}}$ gives a tentative value of $N_{\rm h}$(IP)=0.132 at $\overline{N_{\rm h}}=0.140$, since it cannot be estimated directly from the $K_{\rm s}(T)$ at IP at $y=1$ (see an open circle in Fig.\ref{fig:av} and the text).}
\label{t:ggg}
\end{table}

\begin{figure}[htpb]
\begin{center}
\includegraphics[scale = 0.4]{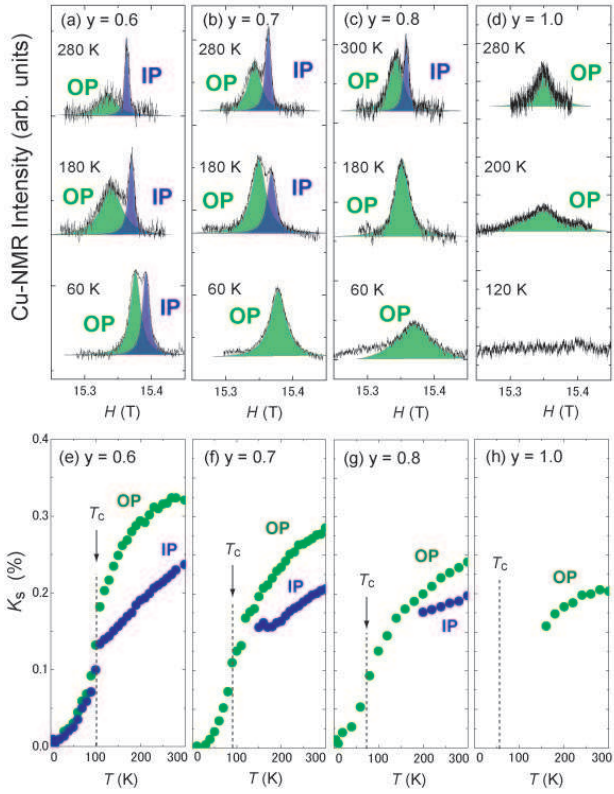}
\end{center}
\caption{\footnotesize (color online) $^{63}$Cu-NMR spectra of the central transition (1/2 $\Leftrightarrow$ $-$1/2) for (a) $y=0.6$, (b) $y=0.7$, (c) $y=0.8$, and (d) $y=1.0$. The temperature dependence of the Knight shift with $H$ perpendicular to the $c$-axis for (e) $y=0.6$, (f) $y=0.7$, (g) $y=0.8$, and (h) $y=1.0$.}
\label{fig:KS}
\end{figure}

Figure \ref{fig:KS} indicates typical $^{63}$Cu-NMR spectra of the central transition (1/2 $\Leftrightarrow$ $-$1/2) for (a) $y=0.6$, (b) $y=0.7$, (c) $y=0.8$, and (d) $y=1.0$. The field-swept NMR spectra were measured at $H$ perpendicular to the $c$-axis. The NMR spectral widths for the samples at room temperature are as narrow as that of Hg-1245,\cite{Kotegawa2004,Mukuda2008} assuring the high quality of the samples. The Cu-NMR spectra for OP and IP are separately detected at $y=0.6$, whereas the IP's spectra for $y=0.7$ and 0.8 disappear at low temperatures due to the development of AFM correlations upon cooling as well as the case for the five-layered Hg- or Tl-based compounds.\cite{Kotegawa2004,Mukuda2008} Surprisingly, the NMR spectrum of OP at $y=1.0$ also disappears at low temperatures, as shown in Fig.\ref{fig:KS}(d), suggesting an onset of AFM order at the OP. The systematic variation of the NMR spectra indicates that Ba$_2$Ca$_3$Cu$_4$O$_8$(F$_y$O$_{1-y}$)$_2$ becomes underdoped as the nominal fluorine content $y$ increases, which ensures that the substitution of oxygen for the apical fluorine increases $N_{\rm h}$. 

\begin{figure}[htpb]
\begin{center}
\includegraphics[scale = 0.3]{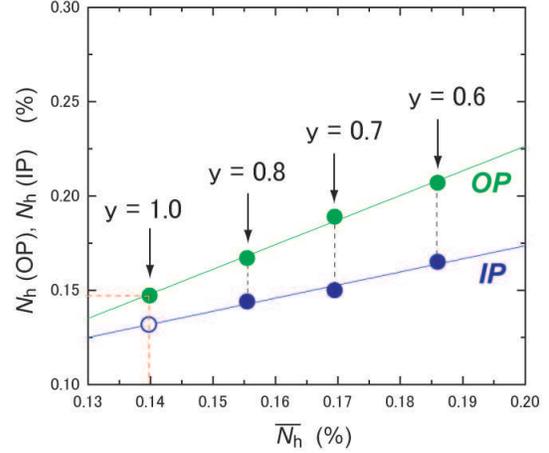}
\end{center}
\caption{\footnotesize (color online)  Plots of $N_{\rm h}$(OP) and $N_{\rm h}$(IP) against averaged carrier density $\overline{N_{\rm h}}$, defined as ($N_{\rm h}({\rm OP})$+$N_{\rm h}({\rm IP})$)/2. Here $N_{\rm h}$(OP) and $N_{\rm h}$(IP) are determined from the Knight shift measurement (see text). $\overline{N_{\rm h}}=0.140$ at $y$=1.0 is estimated from $N_{\rm h}$(OP) = 0.148 on a linear line in the plot of $N_{\rm h}$(OP) versus $\overline{N_{\rm h}}$. A linear extrapolation in the  plots in $N_{\rm h}$(IP) versus $\overline{N_{\rm h}}$ gives a tentative value of $N_{\rm h}$(IP) = 0.132 at $\overline{N_{\rm h}}=0.140$.}
\label{fig:av}
\end{figure}

In general, the Knight shift ($K$) comprises the $T$-dependent spin part ($K_{\rm s}(T)$) and the $T$-independent orbital part ($K_{\rm orb}$); $K=K_{\rm s}(T)+K_{\rm orb}$. Here, $K_{\rm orb}$ was determined as 0.23 ($\pm0.02$)\%, assuming $K_{s}\approx 0$ in the $T=$ 0 limit. The $T$ dependences of $K_{\rm s}(T)$ with $H$ perpendicular to the $c$-axis are displayed in Figs.\ref{fig:KS}(e),(f),(g), and (h) for $y=0.6$, 0.7, 0.8, and 1.0, respectively. The $K_{\rm s}(T)$ decreases upon cooling down to $T_{\rm c}$ for all samples in association with an opening of pseudogap \cite{Yasuoka,REbook}, whereas its steep decrease below $T_{\rm c}$ evidences the reduction of spin susceptibility due to the formation of spin-singlet pairing.  We note that the empirical relation between $K_{\rm s}(T)$ at room temperature and the $N_{\rm h}$ in a CuO$_2$ plane \cite{Zheng,Tokunaga_JLTP} allows us to evaluate $N_{\rm h}$s at OP and IP for the four samples, which are summarized in Table \ref{t:ggg} along with the $T_{\rm c}$ and the averaged carrier density $\overline{N_{\rm h}}$, defined as ($N_{\rm h}$(OP) + $N_{\rm h}$(IP))/2.
Figure \ref{fig:av} indicates the plot of $N_{\rm h}$s at IP and OP against $\overline{N_{\rm h}}$. 
The $N_{\rm h}$(IP) at $y=1.0$, however, was not directly estimated from the $K_{\rm s}(T)$, because the Cu-NMR spectrum was not detected at room temperature. Instead, $\overline{N_{\rm h}}=0.140$ at $y$ = 1.0 is estimated from $N_{\rm h}$(OP)=0.148 on a linear line in the plot of $N_{\rm h}$(OP) versus $\overline{N_{\rm h}}$ in Fig.\ref{fig:av}.  Furthermore, a linear extrapolation in the  plot of $N_{\rm h}$(IP) versus $\overline{N_{\rm h}}$ gives a tentative value of $N_{\rm h}$(IP)=0.132 at $\overline{N_{\rm h}}=0.140$.  As summarized in Table \ref{t:ggg}, the increase of $N_{\rm h}$(OP) and  $N_{\rm h}$(IP) with increasing a nominal oxygen content at the apical site increases $T_{\rm c}$ from 55 K to 102 K.


\begin{figure}[htpb]
\begin{center}
\includegraphics[scale = 0.4]{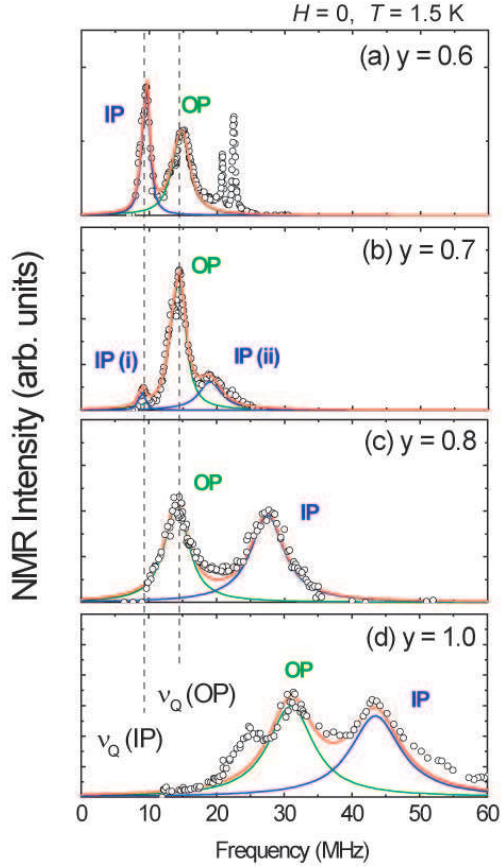}
\end{center}
\caption{\footnotesize (color online) Cu-NQR/ZFNMR spectra at $H=0$ and $T=$1.5 K. (a) The respective Cu-NQR spectra at IP and OP for $y=0.6$ with $^{63}\nu_{\rm Q}$s = 9.7 and 15 MHz. Two sharp peaks at 20$-$25 MHz arise from unknown impurity phases. (b) The Cu-NQR/ZFNMR spectra at $y=0.7$. The NQR spectrum for the OP is observed with almost the same $^{63}\nu_{\rm Q}$ as that of the OP at $y=0.6$. The NQR spectrum at IP(i) is observed at 9.1 MHz that is close to $^{63}\nu_{\rm Q}=9.7$ MHz of the IP at $y=0.6$, whereas the ZFNMR spectrum at IP(ii) probes the internal field $H_{\rm int}$ = 2.4 T due to the AFM order. The fact that the IP(i) and IP(ii) originate from a paramagnetic phase and an AFM phase, respectively suggests that the phase separation takes place because of the closeness to the QCP at which the AFM collapses. (c) The Cu-NQR/ZFNMR spectra at $y=0.8$. The NQR spectrum for the OP is observed at almost the same frequency as that of the OP at $y=0.6$, revealing that no spontaneous moment is induced at low temperatures. The spectrum observed around 28 MHz arises from the IP with $H_{\rm int}\sim$ 2.4 T and hence the AFM moment $M_{\rm AFM}\sim0.12$ $\mu_{\rm B}$. (d) The Cu-NMR spectra at $y=1.0$ with $H_{\rm int}\sim$ 2.7 T and 3.8 T for the OP and IP, respectively. Here, $M_{\rm AFM}=0.11$ $\mu_{\rm B}$ and 0.18 $\mu_{\rm B}$ are estimated for the OP and IP, respectively.}
\label{fig:zero}
\end{figure}
\subsection{Zero-field NMR evidence of AFM order}

We deal with the AFM order taking place in the uderdoped CuO$_2$ layers. The observation of zero-field NMR (ZFNMR) spectra enables us to assure an onset of an AFM order, since magnetically ordered moments induce internal magnetic field $H_{\rm int}$ at nuclear sites. Generally, the Hamiltonian for Cu nuclear spin with $I=3/2$ is described by the Zeeman interaction due to magnetic field $H$ (${\cal H}_{\rm Z}$) and the nuclear-quadrupole interaction (${\cal H}_{\rm Q}$) as follows:

\begin{equation}
{\cal H}={\cal H_{\rm Z}}+{\cal H_{\rm Q}}=-\gamma_{\rm N}\hbar {\bm I} \cdot {\bm H}+\frac{e^{2}qQ}{4I(2I-1)}(3I_{\rm z^{\prime}}^2-I(I+1)),
\label{eq:hamiltonian}
\end{equation}
where $\gamma_{\rm N}$ is the Cu nuclear gyromagnetic ratio, $eQ$ is the nuclear quadrupole moment, and $eq$ is the electric field gradient (EFG) at the Cu nuclear site. Here, in the ${\cal H}_{\rm Q}$, an asymmetric parameter ($\eta$) is zero in the tetragonal symmetry. Note that the nuclear quadrupole resonance (NQR) frequency $\nu_{\rm Q}=3e^{2}qQ/2h I(2I-1)$. The nuclear Hamiltonian given by eq.(\ref{eq:hamiltonian}) is described with $H_{\rm int}$ instead of $H$ for zero-field experiments.

Figure \ref{fig:zero}(a) indicates the Cu-NQR spectrum at $y=0.6$. Respective $^{63}\nu_{\rm Q}$s are evaluated as 9.7 and 15 MHz at the IP and OP, which are comparable to $\sim$ 8$-$10 MHz and $\sim$ 16 MHz for five-layered systems \cite{Tokunaga,Kotegawa2001,Kotegawa2004}. Here, the sharp $^{63}$Cu- and $^{65}$Cu-NQR spectral widths at $^{63}\nu_{\rm Q}=$ 22.5 and $^{65}\nu_{\rm Q}=$ 20.8 MHz are as narrow as about 400 kHz. These NQR spectra qualitatively differ from those reported  for various copper oxides.\cite{Ohsugi_JLTP, Ohsugi, Ishida_YBCO, Zheng_Tl}  Integrated intensities of these NQR spectra are an order of magnitude smaller than those for the intrinsic phase, which suggests that these NQR spectra arise from some impurity phases containing copper such as starting materials prepared for the sample synthesis, and intermediate products in the high-pressure synthesis.\cite{Iyo2, Iyo1, Takano}

Figure \ref{fig:zero}(b) indicates the Cu-NQR/ZFNMR spectra at $y=0.7$. The NQR spectrum for OP is observed with almost the same $^{63}\nu_{\rm Q}$ as that of the OP at $y=0.6$.  Note that the respective NQR and ZFNMR spectra at IP(i) and IP(ii) arise from IP. The NQR spectrum at IP(i) is observed at 9.1 MHz that is close to $^{63}\nu_{\rm Q}=9.7$ MHz of the IP at $y=0.6$, whereas the ZFNMR spectrum at IP(ii) is observed at $\sim$ 18 MHz. Assuming $^{63}\nu_{\rm Q} = 9.1$ MHz, $H_{\rm int}\sim$1.5 T is estimated for the IP(ii). The $H_{\rm int}$ at CuO$_2$ plane is generally given by $H_{\rm int}=|A_{\rm hf}|M_{\rm AFM}=|A-4B|M_{\rm AFM}$, where $A$ and $B$ are the on-site hyperfine field and the supertransferred hyperfine field from the four nearest neighboring Cu-AFM moments, respectively, and $M_{\rm AFM}$ is the AFM moment.\cite{MilaRice}
Here $A\sim$ 3.7 T/$\mu_{\rm B}$, $B({\rm OP})\sim$ 7.4 T/$\mu_{\rm B}$, and $B({\rm IP})\sim$ 6.1 T/$\mu_{\rm B}$ are assumed to be the same as those for Hg-1245.\cite{Kotegawa2004} Using these values, a uniform AFM moment at the IP(ii) is estimated at $M_{\rm AFM}$(IP) $\sim$ 0.08 $\mu_{\rm B}$ for an AFM phase at $y=0.7$.
The fact that IP(i) and IP(ii) originate from a paramagnetic phase and an AFM phase, respectively, suggests that the phase separation takes place because of the closeness to the QCP at which the AFM collapses.  The presence of the phase separation probably imply that the AFM critical point could be close to $N_{\rm h}$ $\sim$ 0.15. 

Figure \ref{fig:zero}(c) indicates the Cu-NQR/ZFNMR spectra at $y=0.8$. A spectrum observed around 14.4 MHz arises from OP since its peak frequency is almost the same frequency as $^{63}\nu_{\rm Q}$ = 15 MHz for the OP at $y=0.6$. Accordingly, another spectrum around 28 MHz is assigned to arise from IP. Using above-mentioned parameters, $H_{\rm int}\sim$ 2.4 T and $M_{\rm AFM}$(IP) $\sim$ 0.12 $\mu_{\rm B}$ are estimated for the IP at $y=0.8$.

Figure \ref{fig:zero}(d) indicates the Cu-ZFNMR spectra observed around 30 and 45 MHz at $H=0$ for $y=1.0$. When noting that ZFNMR spectra are absent around $^{63}\nu_{\rm Q}$(IP) = 8 $\sim$ 10 MHz and $^{63}\nu_{\rm Q}$(OP) = 14 $\sim$ 16 MHz, the observation of the NMR spectra around 30 MHz and 45 MHz demonstrates that $H_{\rm int}$s are present at the respective IP and OP with $H_{\rm int}\sim$ 3.8 T and 2.7 T. Since $N_{\rm h}$(OP) $>$ $N_{\rm h}$(IP) due to the charge imbalance between OP and IP and hence $M_{\rm AFM}$(OP) $<$ $M_{\rm AFM}$(IP), $M_{\rm AFM}$(OP) $\sim$ 0.11 and  $M_{\rm AFM}$(IP)$\sim$0.18 $\mu_{\rm B}$ are evaluated at the OP and IP, respectively, using the relation of $H_{\rm int}=|A_{\rm hf}|M_{\rm AFM}=|A-4B|M_{\rm AFM}$. Notably, the OP, which is mainly responsible for the SC with $T_{\rm c}$ = 55 K, manifests the AFM order, leading us to a conclusion that the uniform mixing of AFM with $M_{\rm AFM}=0.11$ $\mu_{\rm B}$ and SC at $T_{\rm c}$ = 55 K occurs in the OP as well as in the three IPs of the five-layered systems.\cite{Mukuda2008}  


\begin{figure}[h]
\begin{center}
\includegraphics[scale = 0.4]{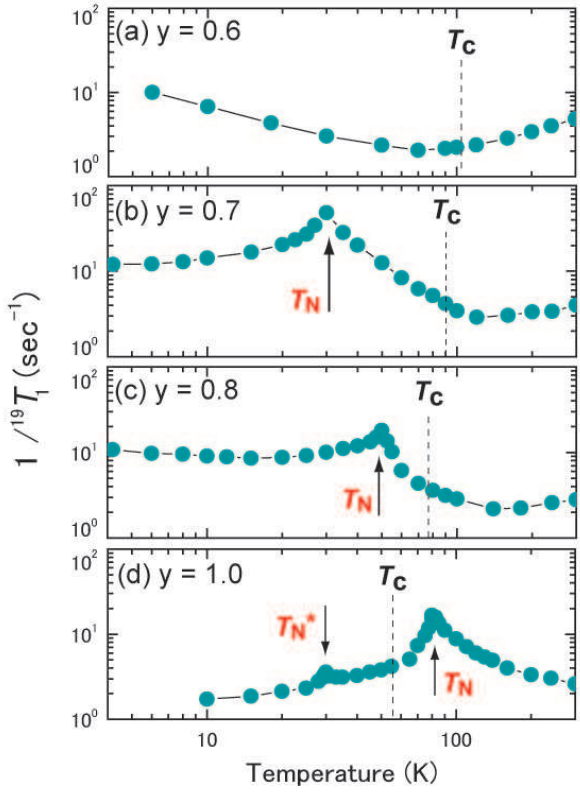}
\end{center}
\caption{\footnotesize (color online) The temperature dependences of $^{19}$($1/T_1$)s for (a) $y=0.6$, (b) $y=0.7$, (c) $y=0.8$, and (d) $y=1.0$ at an NMR frequency 174.2 MHz and $H$ parallel to the $c$-axis. For (a) $y=0.6$, a reason that $1/T_1$ increases upon cooling is associated with the development of magnetic correlations at low temperatures. For (b) $y=0.7$ and (c) $y=0.8$, $1/T_1$ exhibits a peak at $T_{\rm N}\sim$ 30 K and $T_{\rm N}\sim$ 50 K, respectively. (d) There are a distinct peak in $1/T_1$ at $T_{\rm N}\sim$ 80 K and a significant one at $T_{\rm N}^{*}\sim$ 30 K. The AFM order inherent to the OP responsible for SC is presumably developed below $T_{\rm N}^{*}$, exhibiting the spontaneous AFM moment of $M_{\rm AFM}$(OP) $\sim$ 0.11 $\mu_{\rm B}$ at low temperatures.}
\label{fig:TN}
\end{figure}

\subsection{$^{19}$F-NMR probe of N\'eel temperature}

N\'eel temperatures ($T_{\rm N}$s) at $y=0.7$, 0.8 and 1.0 are determined by the $^{19}$F-$T_{\rm 1}$ measurement, which sensitively probes critical magnetic fluctuations developing at OP and IP as the system approaches an AFM order. Generally, $1/T_{\rm 1}$ is described as follow;
\begin{equation}
\frac{1}{T_{1}}=\frac{2\gamma_{\rm N}^{2}k_{\rm B}T}{(\gamma_{\rm e} \hbar )^{2}}\sum_{\bm q} |A_{\bm q}|^{2}\frac{{\rm Im}[\chi(\bm q, \omega_{0})]}{\omega_{0}},
\label{eq:T1}
\end{equation}
where $A_{\bm q}$ is the wave-vector (${\bm q}$)-dependent hyperfine-coupling constant, $\chi({\bm q},\omega)$ is the dynamical spin susceptibility, and $\omega_0$ is the NMR frequency.  The $T$ dependence of $1/T_1$ shows a peak at $T_{\rm N}$ because the low-energy spectral weight in $\chi({\bm q}={\bm Q},\omega)$ is strongly enhanced around $\omega_0 \sim 0$ in association with a divergence of magnetic correlation length at $T \sim T_{\rm N}$.  Here ${\bm Q}$ is the AFM wave vector ($\pi/a$, $\pi/a$). $^{19}(1/T_{\rm 1})$ for all samples are presented in Fig.\ref{fig:TN}. In the present case, the relaxation processes in $1/T_1$ compose of quasiparticle contributions probing the onset of SC and the magnetic one probing magnetic fluctuations. However, we consider that the former is negligible in the case of 1/$T_1$ at the apical site; in fact, it has been reported $1/T_1$ at the apical oxygen by $^{17}$O-NMR did not change drastically at $T_{\rm c}$ \cite{Hammel} because of the very small hyperfine-coupling constant with the quasiparticles in CuO$_2$ layers. In this context, $^{19}(1/T_1)$ is expected to be dominated by magnetic fluctuations. 

Respective figures \ref{fig:TN}(b) and (c) show the $T$ dependences of $^{19}(1/T_{1})$s at $y=0.7$ and 0.8, exhibiting the peaks at $\sim$ 30 K and 50 K. This result ensures the AFM order at $T_{\rm N}$ = 30 and 50 K for the IP(ii) at $y=0.7$ and the IP at $y=0.8$ with the spontaneous AFM moment of $M_{\rm AFM}$(IP(ii)) $\sim$ 0.08 $\mu_{\rm B}$ and $M_{\rm AFM}$(IP) $\sim$ 0.12 $\mu_{\rm B}$, respectively. Note that the absence of a peak in $1/T_1$ at $y=0.6$ evidences that this compound is in a paramagnetic state down to 4.2 K. As shown in Fig.\ref{fig:TN}(d) for $y$ = 1.0, there are a distinct peak in $1/T_1$ at $T_{\rm N}$ $\sim$ 80 K and a significant one at $T_{\rm N}^{*}\sim$ 30 K. 
The AFM order inherent to the OP responsible for SC is presumably developed below $T_{\rm N}^{*}\sim$ 30 K, exhibiting the spontaneous AFM moment of $M_{\rm AFM}$(OP) $\sim$ 0.11 $\mu_{\rm B}$ at low temperatures. This suggests that the SC uniformly coexists with the AFM order in  a single CuO$_2$ plane with $N_{\rm h}$ $\sim$ 0.148. 
Since $N_{\rm h}$(IP) $<$ $N_{\rm h}$(OP)  and $M_{\rm AFM}$(IP) $>$ $M_{\rm AFM}$(OP), the $T_{\rm N}$ at the IP becomes larger than at the OP. It is noteworthy that $T_{\rm N}^{*}\sim$ 30 K for the OP at $y$=1.0 is comparable to the $T_{\rm N}\sim$ 30 K for the IP(ii) at $y=$ 0.7 because both layers possess almost the same $N_{\rm h}$.

\subsection{Phase diagram of AFM and SC}

\begin{figure}[htpb]
\begin{center}
\includegraphics[scale = 0.35]{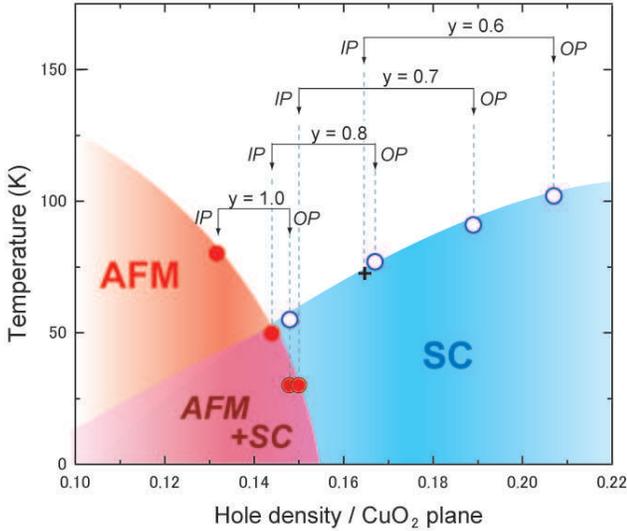}
\end{center}
\caption{\footnotesize (color online) A phase diagram of AFM and SC as a function of hole carrier density $N_{\rm h}$. $T_{\rm c}$ (indicated by open circle) and $T_{\rm N}$ (closed circle) are plotted against $N_{\rm h}$ for the OPs and IPs of Ba$_2$Ca$_3$Cu$_4$O$_8$(F$_y$O$_{1-y}$) at $y$ = 0.6, 0.7, 0.8 and 1.0.  $N_{\rm h}$ was determined from the Knight shift measurement (see the text). The $T_{\rm c}\sim$ 70 K for the IP at $y=$ 0.6, that is shown by  black cross (${\bm +}$) was determined from a small peak at $T\sim$ 70 K in the $T$ derivative of the Knight shift ($dK/dT$) as well as in the literatures.\cite{Tokunaga,Mukuda2008}  Note that the uniform mixing of AFM ($T_{\rm N}$ = 30 K) and SC ($T_{\rm c}$ = 55 K) takes place at the OP at $y$ = 1.0. This result strongly suggests that it is a general property inherent to a single CuO$_2$ plane in the underdoped regime for hole-doping.\cite{Mukuda2006,Mukuda2008}}
\label{fig:PD}
\end{figure}

Figure \ref{fig:PD} reveals a phase diagram of AFM and SC as a function of $N_{\rm h}$ where  $T_{\rm c}$ and $T_{\rm N}$($T_{\rm N}^{*}$) are plotted against $N_{\rm h}$ for the OPs and IPs of the four-layered superconductors Ba$_2$Ca$_3$Cu$_4$O$_8$(F$_y$O$_{1-y}$) at $y$=0.6, 0.7, 0.8, and 1.0. We remark that the uniform mixing of AFM ($T_{\rm N}$ = 30 K) and SC ($T_{\rm c}$ = 55 K) was observed for the OP at $y$ = 1.0, which strongly suggests that it is a general property inherent to a single CuO$_2$ plane in the underdoped regime for hole-doping.  
It has been reported in the literatures \cite{Kotegawa2001,Mukuda2008} that a bulk $T_{\rm c}$ in multilayered compounds was determined by the $T_{\rm c}$ of OP and that the $T_{\rm c}$ of IP was significantly lower than a bulk $T_{\rm c}$ due to the lower $N_{\rm h}$ at IP. In the phase diagram in Fig.\ref{fig:PD}, the QCP in the four-layered system is obtained at $N_{\rm h}$ $\simeq$ 0.15 smaller than $N_{\rm h}$ $\simeq$ 0.17 for the five-layered system \cite{Mukuda2008}, suggesting that the interlayer magnetic coupling of the four-layered compound is smaller than that of the five-layered compound.
The phase diagrams of AFM and SC in multilayered systems are remarkably different from the well-established ones for LSCO ($n=$ 1) \cite{LSCO} and YBCO ($n=$ 2), \cite{YBCO} where the AFM order totally collapses by doping very small amount of holes with $N_{\rm h}$ $\sim$ 0.02 and $N_{\rm h}$ $\sim$ 0.055, respectively.  
The reason that the AFM phase exists up to $N_{\rm h}$ $\simeq$ 0.15 and 0.17 in the four- and five-layered compounds, respectively, is because the interlayer magnetic couplings are stronger than in LSCO or YBCO due to the existence of the homogeneously underdoped IPs.

\section{Summary}
The extensive Cu-NMR/NQR and F-NMR measurements on the four-layered high-$T_{\rm c}$ superconductors Ba$_2$Ca$_3$Cu$_4$O$_8$(F$_y$O$_{1-y}$)  have unraveled the systematic evolution of AFM and SC as the function of hole carrier density $N_{\rm h}$; $T_{\rm c}$ and $T_{\rm N}$ are controlled by the substitution of oxygen for fluorine at the apical site. It is demonstrated that
the AFM order, which can uniformly coexist with SC, exists up to $N_{\rm h}\simeq$ 0.15, reinforcing that the uniform mixing of AFM and SC is a general property inherent to a single
CuO$_2$ plane in the underdoped regime for hole-doping.  $N_{\rm h}\simeq$ 0.15 at QCP for the four-layered compounds is somewhat smaller than $N_{\rm h}$ $\simeq$ 0.17 for the five-layered compounds. The fact that the QCP for the four-layered compounds moves to a region of lower carrier density than for the five-layered compounds ensures that the decrease in the number of CuO$_2$ layers makes an interlayer magnetic coupling weaker. The present studies have highlighted the intimate evolution of AFM and SC in the phase diagram inherent to the homogeneously doped CuO$_2$ plane, which depends on the interlayer magnetic coupling significantly.

\section*{Acknowledgments}

The authors are grateful to M. Mori, T. Tohyama, H. Eisaki and M. Ogata for their helpful discussions. This work was supported by a Grant-in-Aid for Specially Promoted Research (20001004) and in part by Global COE Program (Core Research and Engineering of Advanced Materials Science), from the Ministry of Education, Culture, Sports, Science and Technology (MEXT), Japan. One of the authors (S.S.) is financially supported as a JSPS (the Japan Society for the Promotion of Science) Research Fellow.

\end{document}